\documentclass[11pt,english,superscriptaddress,showpacs,aps]{revtex4}
\usepackage{graphicx}
\usepackage{amssymb}
\usepackage{babel}

\usepackage{hyperref}



\begin{document}

\title{A coherent state approach to effective potential in noncommutative $D=(2+1)$ models}

\author{A. C. Lehum}
\email{andrelehum@ect.ufrn.br}
\affiliation{Escola de Ci\^encias e Tecnologia, Universidade Federal do Rio Grande do Norte\\
Caixa Postal 1524, 59072-970, Natal, RN, Brazil}

\author{J. R. Nascimento}
\email{jroberto@fisica.ufpb.br}
\affiliation{Departamento de F\'{\i}sica, Universidade Federal da Para\'{\i}ba\\
 Caixa Postal 5008, 58051-970, Jo\~ao Pessoa, Para\'{\i}ba, Brazil}

\author{A. Yu. Petrov}
\email{petrov@fisica.ufpb.br}
\affiliation{Departamento de F\'{\i}sica, Universidade Federal da Para\'{\i}ba\\
 Caixa Postal 5008, 58051-970, Jo\~ao Pessoa, Para\'{\i}ba, Brazil}

\begin{abstract}

In this work we study the effective potential in noncommutative three-dimensional models where the noncommutativity is introduced through the coherent state approach. We discuss some important characteristics that seem to be typical to this approach, specially the behavior of the quantum corrections in the small noncommutativity limit.  

\end{abstract}

\maketitle

The noncommutativity of spacetime coordinates characterized by their following commutation relation, 
\begin{eqnarray}\label{eq0}
[\hat{q}_1^\mu,\hat{q}_2^\nu]=i\Theta^{\mu\nu}~,
\end{eqnarray}

\noindent
was originally proposed~\cite{Snyder:1946qz} as a way to avoid the ultraviolet (UV) divergences which arise within the perturbative approach to quantum field theory. Many years later, it was discovered that in a certain low energy limit of the string theory, a noncommutative field theory (NCFT) emerges~\cite{Seiberg:1999vs}. This result inspired the scientific community to investigate different aspects of the NCFTs, while one of the most intriguing issues is the UV/IR (ultraviolet/infrared) mixing~\cite{Minwalla:1999px}, that is the mixing of very distinct energy scales implying in arising of the new, so-called UV/IR infrared singularities, which can destroy the perturbative expansion. Other important issue is the lack of unitarity in theories where time does not commute with space coordinates~\cite{Gomis:2000zz}. This can be avoided via consideration of the theories with only space-space noncommutativity. Another observation is that the relation Eq.(\ref{eq0}) is not Lorentz invariant, unless that we promote the noncommutative parameter $\Theta^{\mu\nu}$ to be an operator~\cite{Bahns:2002vm,Doplicher:1994tu}, in contrast with the constant noncommutativity. One should observe that the noncommutativity of spacetime coordinates, in general, does not exclude the possibility for the appearance of UV divergences and a renormalization prescription is still necessary to guarantee the consistence of NCFT. 

Recently, one of the subjections to implement the noncommutativity of the spacetime coordinates, that is the coherent states formalism~\cite{Smailagic:2003rp,Smailagic:2003yb,Smailagic:2004yy}, have attracted some attention because it apparently solves the problems with respect to unitarity, Lorentz invariance, and even the finiteness problem. The finiteness is achieved because the coherent states approach to NCFT introduces a natural cutoff $\Theta$, that is the parameter of the noncommutativity, in the propagators of the model making any Feynman amplitude to be finite, at least while $\Theta$ is nonvanishing. In the three-dimensional case that we consider here, the coherent states naturally emerge in the context of anyons~\cite{anyons}.

One remarkable result is the renormalization of $D=(2+1)$ Gross-Neveau model, whereas the only consistent noncommutative extension of this model was obtained through coherent state approach~\cite{Charneski:2006be}. A large number of applications of this formalism have been studied, as such Aharonov-Bohm scattering~\cite{Anacleto:2005mq} and some black holes effects~\cite{Nicolini:2005vd,Ansoldi:2006vg}.

In this work, using the tadpole method~\cite{Weinberg:1973ua}, we will study the first quantum correction to the effective potential in a noncommutative three-dimensional spacetime based on the coherent state formalism, discussing some issues and difficulties of interpretation that was found in our approach.   

In the coherent state formalism the commutation relation Eq.(\ref{eq0}) between the coordinates $q_1$ and $q_2$ implies that the variable $\hat{z}=\frac{1}{\sqrt{2}}(\hat{q}_1+i\hat{q}_2)$ and its complex conjugate $\hat{z}^{\dagger}=\frac{1}{\sqrt{2}}(\hat{q}_1+i\hat{q}_2)$ satisfy the following commutation relation 
\begin{eqnarray}\label{eq0a}
[\hat{z},\hat{z}^\dagger]=\Theta~,
\end{eqnarray}

\noindent
from which we can define a vacuum state $|0\rangle$ as the state satisfying the relation $\hat{z}|0\rangle=0$~\cite{Smailagic:2003rp,Smailagic:2003yb,Smailagic:2004yy}.

A coherent state is defined as an eigenstate of the annihilation operator $\hat{z}$ given by
\begin{eqnarray}\label{eq0b}
|\alpha\rangle=\exp\left({-\frac{|\alpha|^2}{2}\alpha \hat{z}^\dagger}\right)|0\rangle,
\end{eqnarray}

\noindent
where $\hat{z}|\alpha\rangle=\alpha|\alpha\rangle$.

If $\alpha=x+iy$, where $x$ and $y$ are commutative coordinates, a classical field $\phi(x)$ will be represented by 
\begin{eqnarray}\label{eq0c}
\phi(x)=\langle\alpha|\phi(\hat{q})|\alpha\rangle=\int\frac{d^3k}{(2\pi)^3}e^{-ikx-\frac{\Theta}{4}|\vec{k}|^2} \tilde{\phi}(k)~,
\end{eqnarray}

\noindent
where $\tilde{\phi}(k)$ denotes the Fourier transform of $\phi(x)$. 

Thus, with the coherent state representation of the classical field $\phi(x)$, we can determine its propagator which turns out to be
\begin{eqnarray}\label{eq0d}
\Delta(x-y)&\equiv& \langle0|T~\phi(x)\phi(y)|0\rangle\nonumber\\
&=&\int\frac{d^3k}{(2\pi)^3}e^{-ik(x-y)}\frac{e^{-\frac{\Theta}{2}|\vec{k}|^2}}{k^2+m^2}~,
\end{eqnarray}

\noindent
where the chosen signature is $(-++)$.

From the equation above, it is easy to see that the free Lagrangian of the theory within the coherent states approach can be formally represented in a similar form to~\cite{Charneski:2006be},
\begin{eqnarray}\label{eq0e}
\mathcal{L}=\frac{1}{2}\phi(x)(\Box-m^2)e^{-\frac{\Theta}{2}\nabla^2}\phi(x)~.
\end{eqnarray}

Now, let us develop the perturbative approach to the quantum field theories within the coherent states formalism. As a first example, let us consider a very simple model defined by the action
\begin{eqnarray}\label{eqa1}
S&=&\int{d^3x}\Big\{\frac{1}{2}\phi\Box e^{-\frac{\Theta}{2}\nabla^2}\phi -8g^2\phi^6\Big\}~,
\end{eqnarray}

\noindent
which describes a massless real self-interacting scalar field.

To evaluate the effective potential, let us dislocate the field $\phi$ by $\Phi$, $\phi\rightarrow(\phi+\Phi)$, where $\Phi$ is interpreted as the classical background field, and $\phi$ is a quantum one (cf. \cite{BO}). So, the action (\ref{eqa1}) can be rewritten in terms of new fields as
\begin{eqnarray}\label{eqa2}
S&=&\int{d^3x}\Big\{\frac{1}{2}\phi\left(\Box-M^2_\phi\right)e^{-\frac{\Theta}{2}\nabla^2}\phi 
-8g^2\phi^6 -48g^2\Phi\phi^5\nonumber\\
&& -120g^2\Phi^2\phi^4 -160g^2\Phi^3\phi^3 - 48g^2\Phi^5\phi-8g^2\Phi^6\Big\}.
\end{eqnarray}

\noindent
where $M^2_\phi=240g^2\Phi^4$ is a background dependent mass.
 
\begin{figure}[ht]
\includegraphics[width=6cm]{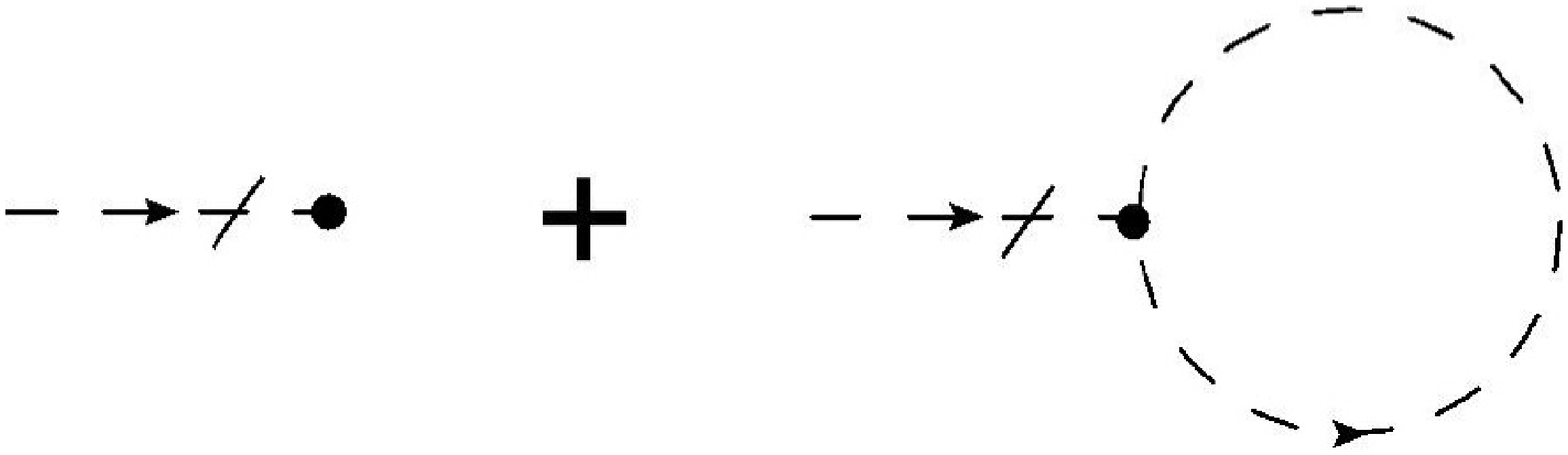}
\caption{One loop tadpole equation. Dashed lines represent the scalar field $\phi$ propagator. Cut lines represent a removed external propagator.}\label{gapsf}
\end{figure}

We will evaluate the effective potential with use of the tadpole method~\cite{Weinberg:1973ua}. The diagram that contributes to tadpole equation at the one-loop order is depicted in Figure \ref{gapsf}, and the corresponding expression can be cast as
\begin{eqnarray}\label{eqa3}
\Gamma^{(1)}_{1}&=&-48ig^2\Phi^5 +480g^2\Phi^3\int \frac{d^3k}{(2\pi)^3} \frac{e^{-\frac{\Theta}{2}|\vec{k}|^2}}{k^2+M^2_\phi}\nonumber\\
&=&-48ig^2\Phi^5-120ig^2\Phi^3 \frac{e^{\frac{\Theta}{2}M^2_\phi}}{\sqrt{2\pi\Theta}} Erfc\left[M_\phi\sqrt{\frac{\Theta}{2}}\right],
\end{eqnarray}

\noindent
where $Erfc(z)=\displaystyle{\frac{2}{\sqrt{\pi}}\int_z^\infty e^{-t^2}{dt}}$ denotes the complementary error function.

Thus, the effective potential looks like
\begin{eqnarray}\label{eqa4}
V_{eff}&=&i\int{d\Phi}~\Gamma^{(1)}_{1}=8g^2\Phi^6+\frac{120g^2}{\sqrt{2\pi\Theta}}\int{d\Phi}~ \Phi^3~e^{\frac{\Theta}{2}M^2_\phi}~Erfc\left[M_\phi\sqrt{\frac{\Theta}{2}}\right]\nonumber\\
&=&8g^2\Phi^6+\frac{\sqrt{15}g}{\pi\Theta}\Phi^2
+\frac{e^{120g^2\Phi^4\Theta}}{4\Phi^2\sqrt{2\pi\Theta^3}}
\Big\{ \Phi^2-Erf\left[\sqrt{120\Theta}g\Phi^2\right]\Big\}~,
\end{eqnarray}

\noindent
where $Erf(z)=\displaystyle{\frac{2}{\sqrt{\pi}}\int_0^z e^{-t^2}{dt}}=1-Erfc(z)$ is the error function.

Taking into account the asymptotic of the error function at the small argument:
$$
Erf(x)\simeq \frac{2}{\sqrt{\pi}}(x-\frac{x^3}{3}+O(x^5)),
$$
we find that term $\Phi^2$ completely disappears from the effective potential, which assumes its minimal value at $\Phi=0$, so no spontaneous symmetry breaking is detected at one-loop order, just as it seems to happen in three-dimensional models which exhibit conformal invariance at the classical level~\cite{Tan:1996kz,Tan:1997ew,Dias:2003pw,Lehum:2008vn}. 

As we suggest that the noncommutativity parameter $\Theta$ is very small, of the order of square of the Planck length, we can expand $V_{eff}$ around $\Theta=0$ and obtain 
\begin{eqnarray}\label{eqa5}
V_{eff}&=&\frac{1}{4\sqrt{2\pi\Theta^3}}+\frac{30g^2}{\sqrt{2\pi\Theta}}\Phi^4
+8g^2\left(1-\frac{10\sqrt{15}}{\pi}g\right)\Phi^6
+900\sqrt{\frac{2\Theta}{\pi}}g^4\Phi^8+\mathcal{O}[\Theta]~.
\end{eqnarray}

We can see that the one-loop effective potential (\ref{eqa5}), obtained with use of the one-loop tadpole equation (\ref{eqa3}), is finite while $\Theta$ differs from zero. The vacuum energy was shifted by $(4\sqrt{2\pi\Theta^3})^{-1}$, but this is not effectively a problem, because we can redefine $V_{eff}$ to eliminate it. It is important to notice that for very small $\Theta$, $\Theta\ll g^2$, we find that the coupling $\frac{15g^2}{\sqrt{2\pi\Theta}}$ in the second term in Eq.(\ref{eqa5}) is much larger than 1, disabling a perturbative analysis of this effective potential, therefore we cannot at least trust that our result is a good quantum approximation for the classical potential. This result can be naturally treated as an analog of the UV/IR mixing that haunts NCFT (cf. \cite{Anacleto:2005mq} where the similar singularities arose within the coherent states approach to the Aharonov-Bohm effect). While this approach to NCFT keeps the Feynman amplitudes to be finite through a ``natural'' cutoff, some remarkable aspects call the attention, those are, first, a very large vacuum energy $(4\sqrt{2\pi\Theta^3})^{-1}$ and, second, a very large coupling constant arising due to the first quantum corrections. 

It is well-known that supersymmetry improves the UV behavior of ordinary theories, and noncommutative extensions of supersymmetric models are less dangerous in relation to UV/IR mixing than the extensions of non-supersymmetric ones. Therefore, as vacuum energy in supersymmetric models is known to vanish, see e.g. \cite{BK0}, let us test whether the bad issues of Eq.(\ref{eqa5}) arise in the three-dimensional Wess-Zumino model.  

So, let us consider the three-dimensional ${\cal N}=1$ Wess-Zumino model~\cite{Gates:1983nr} whose action being modified via the Gaussian factors within the coherent states approach is defined by
\begin{eqnarray}\label{eqb1}
S&=&\int{d^3x}\Big\{\frac{1}{2}\phi(\Box-m^2) e^{-\frac{\Theta}{2}\nabla^2}\phi 
+\frac{1}{2}\psi^{a}(i{{\gamma^\mu}_a}^b\partial_\mu+m\delta_a^b) e^{-\frac{\Theta}{2}\nabla^2}\psi_{b}\nonumber\\
&&-8g^2\phi^6
+ 6g \phi^2\psi^{a}\psi_{a}\Big\}~,
\end{eqnarray}

\noindent
where spacetime indices are represented by Greek letters running from $0$ to $2$. Latin letters represent spinor indices assuming values $1$ or $2$.

To proceed with the calculation of the effective potential, let us shift the field $\phi$ by the background field $\Phi$. Therefore, Eq.(\ref{eqb1}) can be written as
\begin{eqnarray}\label{eqb2}
S&=&\int{d^3x}\Big\{\frac{1}{2}\phi\left(\Box-M^2_\phi\right)e^{-\frac{\Theta}{2}\nabla^2}\phi +\frac{1}{2}\psi^{a}\left(i{{\gamma^\mu}_a}^b\partial_\mu+M_\psi\delta^a_b\right)e^{-\frac{\Theta}{2}\nabla^2}\psi_{b} \nonumber\\
&&- 8g^2 \phi^6 - 48g^2\Phi\phi^5 -120g^2\Phi^2\phi^4 
-160g^2\Phi^3\phi^3 + 6g\phi^2\psi^{a}\psi_{a}\nonumber\\
&& - 48g^2\Phi^5\phi
+12g\Phi\phi\psi^{a}\psi_{a}-8g^2\Phi^6\Big\},
\end{eqnarray}

\noindent
where $M^2_\phi=m^2+240g^2\Phi^4$ and $M_\psi=m+12g\Phi^2$. 

\begin{figure}[ht]
\includegraphics[width=8cm]{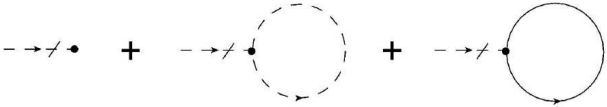}
\caption{Dashed lines represent the scalar field propagator and solid lines represent the fermion field propagator.}\label{gapwz}
\end{figure}

The diagrams which contribute to tadpole equation for the Wess-Zumino model are depicted in Fig. \ref{gapwz}. The bosonic contribution to the tadpole equation is given by
\begin{eqnarray}\label{eqb3s1}
\Gamma^{(1)}_{s}&=&-i\Phi(m^2+48g^2\Phi^4) -120ig^2\Phi^3 \frac{e^{\frac{\Theta}{2}M^2_\phi}}{\sqrt{2\pi\Theta}} Erfc\left[M_\phi\sqrt{\frac{\Theta}{2}}\right],
\end{eqnarray}

\noindent
and the corresponding expression for the fermionic contribution can be cast as
\begin{eqnarray}\label{eqb3}
\Gamma^{(1)}_{f}&=&6ig\Phi M_\psi \frac{e^{\frac{\Theta}{2}M^2_\psi}}{\sqrt{2\pi\Theta}} Erfc\left[M_\psi\sqrt{\frac{\Theta}{2}}\right].
\end{eqnarray}

Integrating the Eqs. (\ref{eqb3s1}) and (\ref{eqb3}) over $\Phi$, and expanding around $\Theta=0$, one find the one-loop effective potential in the form
\begin{eqnarray}\label{eqb6}
V_{eff}&=&\frac{3g\Phi^2(4g\Phi^2-m)}{\sqrt{2\pi\Theta}}+\left(\frac{m^2}{2} +\frac{3gm^2}{\pi}\right)\Phi^2+\frac{m^2}{12\pi}\left(m-\sqrt{m^2+240g^2\Phi^4}\right)
\nonumber\\
&&+\frac{4g^2}{\pi}\Phi^4\left(9m-5\sqrt{m^2+240g^2\Phi^4}\right)
+8g^2\left(1+\frac{18g}{\pi}\right)\Phi^6\nonumber\\
&&-\frac{3g}{2}\sqrt{\frac{\Theta}{2\pi}}\Phi^2(m-4g\Phi^2)(m^2+12gm\Phi^2+192g^2\Phi^4)
+\mathcal{O}[\Theta]~.
\end{eqnarray}

In the supersymmetric theories, the vacuum energy vanishes, i.e., for $\Phi=0$ $V_{eff}=0$. Nevertheless, it is important to notice that the same problem with a large coupling $\frac{12g^2}{\sqrt{2\pi\Theta}}$ still arises. Moreover, if $m\ne0$, the first quantum correction to the tree level mass is given by
\begin{eqnarray}\label{eqb7}
\frac{\partial^2 V_{eff}}{\partial\Phi^2}\Big{|}_{\Phi=0}= m\left[-3g\sqrt{\frac{2}{\Theta\pi}} +m\left(1+\frac{6g}{\pi}\right) +\mathcal{O}[\Theta]^{1/2}\right]~.
\end{eqnarray}

\noindent
Suggesting that $\Theta\sim M_P^{-2}$ ($M_P$ is the Planck mass), one should notice that the correction proportional to $\Theta^{-1/2}$ is negative, therefore a tachyonic state arises when a tree level massive theory is considered unless that $m>M_P$. But, to deal with such massive particles, we should take into account the gravitational effects whose presence would radically modify our study; in particular, the Gaussian factor would be also modified. So, it seems that the appearance of large effective coupling constants is a typical characteristic of the coherent state formulation of NCFTs.  

We could rescale the coupling constant as $g=g'\sqrt{\Theta}$, in this case we can be sure that a large coupling does not arise and no tachyonic state would emerge. But, do we expect that a very weak coupling after quantum corrections could generate a potential $g'\Phi^4$ with the coupling constant $g'$ much larger that $g$? Could this ``miracle'' be a grasping of some unknown low energy quantum gravity effect? A more profound study in other dimensions would be necessary to point a direction. Anyway, these simple models presented here show us some characteristics of coherent state approach to NCFT which were not discussed before.

In this paper, we have calculated the one-loop effective potential in the noncommutative $\phi^6$ theory and its supersymmetric extension formulated on the base of the coherent states approach. The typical feature of theories formulated within this approach are the singularities arising in the small $\Theta$ limit which can in principle imply in the tachyonic instability of the vacuum. Such singularities represent themselves as the natural analog of the infrared singularities arising within the UV/IR mixing mechanism in the usual noncommutative theories based on the Moyal product. We find that the presence of such singularities together with the fact that the dimensional regularization known as a powerful tool implying in removing of some dangerous divergences in the commutative case does not work within the coherent states approach. However, we expect that better supersymmetric extension will allows to rule out such divergences.

\vspace{1cm}
{\bf Acknowledgments.} This work was partially supported by the Brazilian agencies Conselho Nacional de Desenvolvimento Cient\'{\i}fico e Tecnol\'{o}gico (CNPq) and Coordena\c{c}\~ao de Aperfei\c{c}oamento de Pessoal de N\'\i vel Superior (CAPES: AUX-PE-PROCAD 579/2008). The work by A. Yu. P. has been supported by the CNPq project 303461/2009-8.

\newpage

\end{document}